\documentclass[
reprint, 
 amsmath,amssymb,
 aps,
]{revtex4-2}

\usepackage{graphicx}
\usepackage{dcolumn}
\usepackage{url}
\usepackage{bm}
\usepackage{array}

\begin{document}

\preprint{APS/123-QED}

\title{Modeling Insights from COVID-19 Incidence Data: Part I - Comparing COVID-19 Cases Between Different-Sized Populations}

\author{Ryan Wilkinson}
\author{Marcus Roper}

\date{\today}

\begin{abstract}

Comparing how different populations have suffered under COVID-19 is a core part of ongoing investigations into how public policy and social inequalities influence the number of and severity of COVID-19 cases. But COVID-19 incidence can vary multifold from one subpopulation to another, including between neighborhoods of the same city, making comparisons of case rates deceptive. At the same time, although epidemiological heterogeneities are increasingly well-represented in mathematical models of disease spread, fitting these models to real data on case numbers presents a tremendous challenge, as does interpreting the models to answer questions such as: Which public health policies achieve the best outcomes? Which social sacrifices are most worth making? Here we compare COVID-19 case-curves between different US states, by clustering case surges between March 2020 and March 2021 into groups with similar dynamics. We advance the hypothesis that each surge is driven by a subpopulation of COVID-19 contacting individuals, and make detecting the size of that population a step within our clustering algorithm. Clustering reveals that case trajectories in each state conform to one of a small number (4-6) of archetypal dynamics. Our results suggest that while the spread of COVID-19 in different states is heterogeneous, there are underlying universalities in the spread of the disease that may yet be predictable by models with reduced mathematical complexity. These universalities also prove to be surprisingly robust to school closures, which we choose as a common, but high social cost, public health measure.
\end{abstract}

\maketitle

\section{Introduction}

\begin{figure}[!ht]
         \centering
         \includegraphics[scale = 0.3]{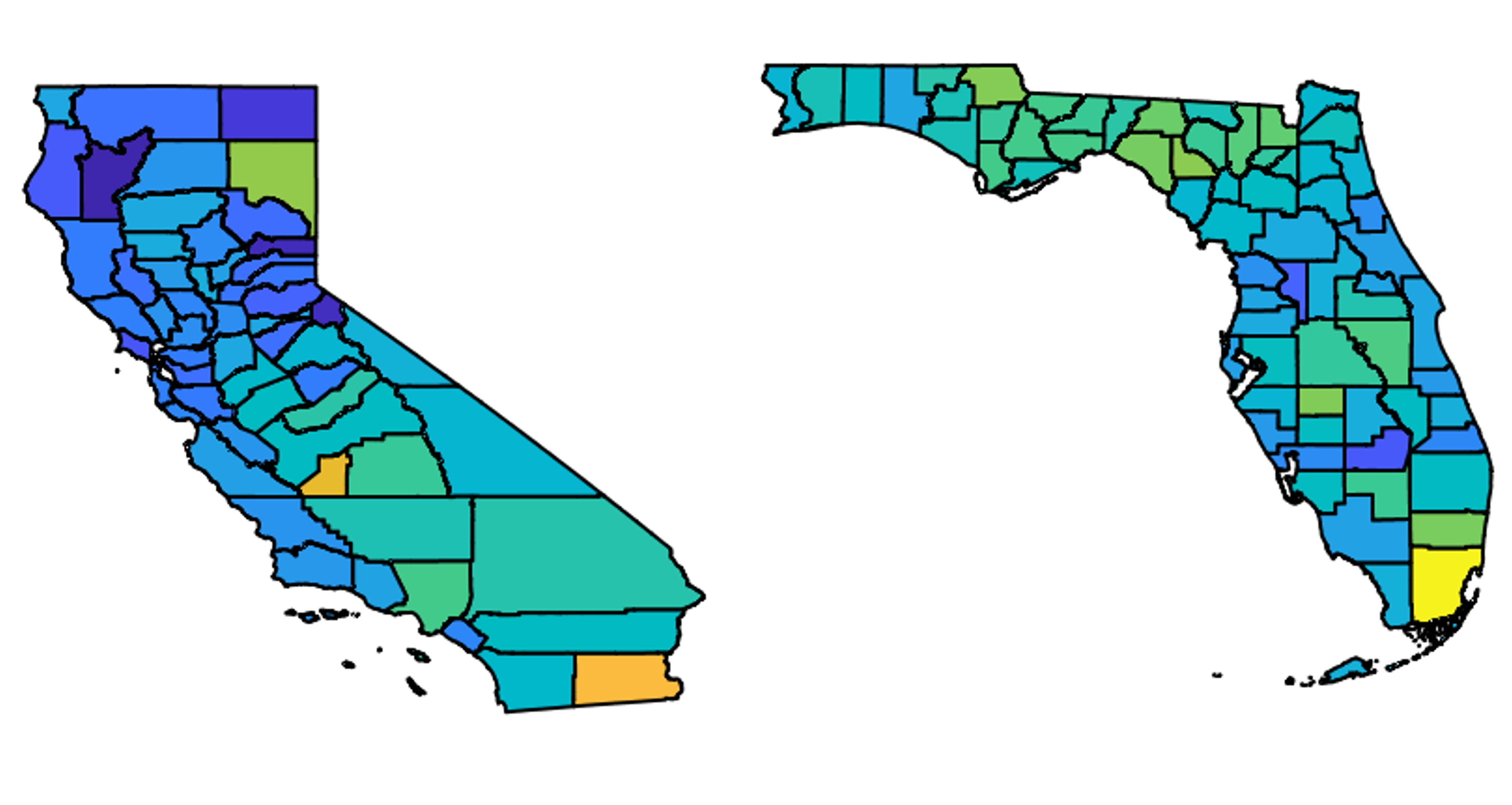}
         \includegraphics[scale = 0.3]{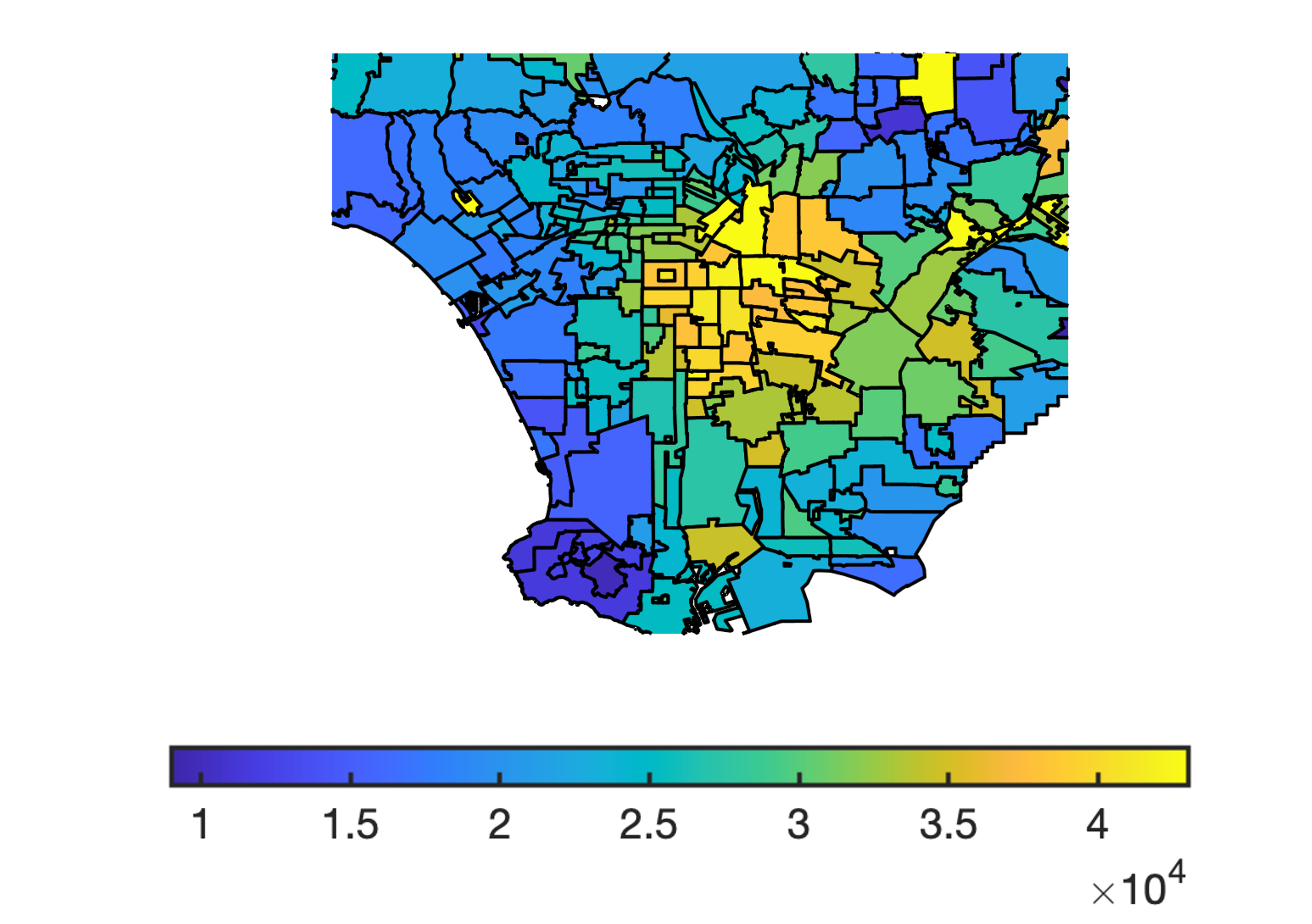}
        \caption{Normalization of cases by total population of state masks small spatial scale heterogeneities in case rates. In California, Kings and Imperial Counties have highest case rates, but numerically largest number of cases is in Los Angeles County (left). Miami-Dade county has highest case numbers and rates in FL (middle). Case rates in Downtown and East Los Angeles County neighborhoods are 2-3 times higher than in West and South Bay neighborhoods (right). Shown: cumulative data number of COVID cases per 100,000 individuals, on April 24, 2022. Sources: FL and CA county data: New York Times, COVID-19 dashboard, LA neigborhood data, Los Angeles Times, COVID-19 dashboard.}
        \label{fig:hetero}
\end{figure}

The different approaches that different US states have taken to controlling or mitigating the spread of COVID-19 have created a test-bed for evaluating public health responses to future diseases. It is natural to ask, and many headline writers have already \cite{woodfolk_compare,browne_compare}, whether specific states have done better or worse than others. California and Florida have been objects of frequent comparison, because of their similar mild winter weather and highly diverse populations, but stark policy differences on school re-openings, masking and indoor dining. However, to compare number of COVID cases, or number of deaths, between different populations, it is necessary to normalize by some measure of population size. Simply dividing by the total state populations, shows slightly larger case rates and much higher death rates in Florida (Table \ref{tab:CA_FL_Table}). By dividing by the total population of the state, we effectively treat the population of the state as a single entity, but both states have highly heterogeneous distributions of cases \ref{fig:hetero}: in Florida both cases and case rates are concentrated in Miami-Dade county, while in California, case rates are highest in two low density rural counties (Imperial County and King's County), though the greatest number of cases occurs again in a single large metropolitan area (Los Angeles). Dividing the total number of cases (dominated by a few urban hotspots) by a total state population leads to a misleading picture of COVID-19 incidence. When comparing COVID incidences between different populations, we would like the population size that we normalize by to reflect a well-mixed subpopulation of individuals with similar levels of exposure and susceptibility to the disease. However, isolating these populations is not straightforward: even we study a single metropolitan area (e.g. Los Angeles county), we find widely different COVID case rates between different neighborhoods, separated only by miles (Fig \ref{fig:hetero}).

\begin{table}
    \centering
    \begin{tabular}{c||c|c}
    & California & Florida \\
    \hline 
    \hline
    Population
    & 39.19 M
    & 21.48 M\\
    \hline
    Cases
    & 9,237,030
    & 5,963,941\\
    \hline
    Deaths
    & 90,706
    & 74,056\\
    \hline
Case rate & 23,570 & 27,380
\\
\hline Death rate & 231 & 340\\ \hline
    \end{tabular}
    \caption{\label{tab:CA_FL_Table} Cumulative numbers of COVID-19 cases and deaths in California and Florida, on April 22nd, 2022. Naive comparisons between states are based on case and death rates /100,000 individuals, calculated by dividing by the total state population. Source: New York Times COVID-19 dashboard. }
\end{table}

Among mathematical models that have been deployed to predict COVID spread, and to assist with the allocation of resources, some, such as agent based and network models, specifically address the role of population heterogeneities in shaping the spread of the disease \cite{keeling2005networks,britton2020mathematical,moreno2002epidemic,arenas2020mathematical,bansal2007individual,bansal2010dynamic,keeling2005implications,li2020simulating,volz2007susceptible}. Other models upscale heterogeneous populations into single, well-mixed, groups of individuals, trading off the flexibility in forming the dynamics of interaction that a more complex model affords, against a smaller set of parameters are easier to interpret and to fit against data \cite{bertozzi2020challenges,sahneh2013generalized,tolles2020modeling,SIR_Mathematics}. A third, influential, class of models is purely data fitted, with no mechanistic interpretation about how individuals are interacting or transmitting the disease \cite{IHME_First_Wave_Forecasting,IHME_Hospital_Forecasting}. 

Here we seek to shed light on both problems: 1. how to compare COVID cases among two populations of two sizes, and 2. what is the appropriate level of complexity to use in a mathematical model describing the spread of COVID-19 within a population.
Specifically, we seek to compare the COVID-19 incidence curves between different US states, as paradigms of large heterogeneous, incompletely mixed populations. Comparing the data reveals that there are natural normalizations for case curves, which we interpret as the size of the subpopulation that has contact with COVID-infected individuals. The second goal of our analysis is to determine how diverse COVID-19 incidence time courses truly are: whether each State follows a completely distinct disease trajectory, or whether there are common families of time courses representing quantitatively similar dynamics of disease spread that could put an upper bound on the complexity of models needed to predict the course of the epidemic. Our third goal is to make trend-conscious comparisons that could be used to quantitatively compare the effectiveness of public health control measures practiced in different populations based on the time course of the pandemic rather than on single point in time measurements, such as those given in Table \ref{tab:CA_FL_Table}. 

To this end, we analyzed one year of cumulative U.S. State case data (from March 2020 to March 2021) taken from The COVID Tracking Project\cite{The_COVID_tracking_project}, and isolated the earliest and last complete waves of spread contained in these data. We clustered case curves of similar shape, normalizing each case curve by a population-size that was allowed to different from the overall state population. We clustered case curves of similar shape together: These clusters revealed surprising case curve homologies that grouped states with very different voting preferences, urban/rural densities, and demographies. Comparisons within clusters gave estimates of the sizes of the populations among which active COVID transmission is occurring, whereas comparisons between clusters may allow for the evaluation of the effectiveness of public health measures controlling spread.

\section{Results}\label{sec:Results}

\subsection{U.S. State Clustering}

\begin{figure*}[!ht]
    \centering
    \includegraphics[scale = .65]{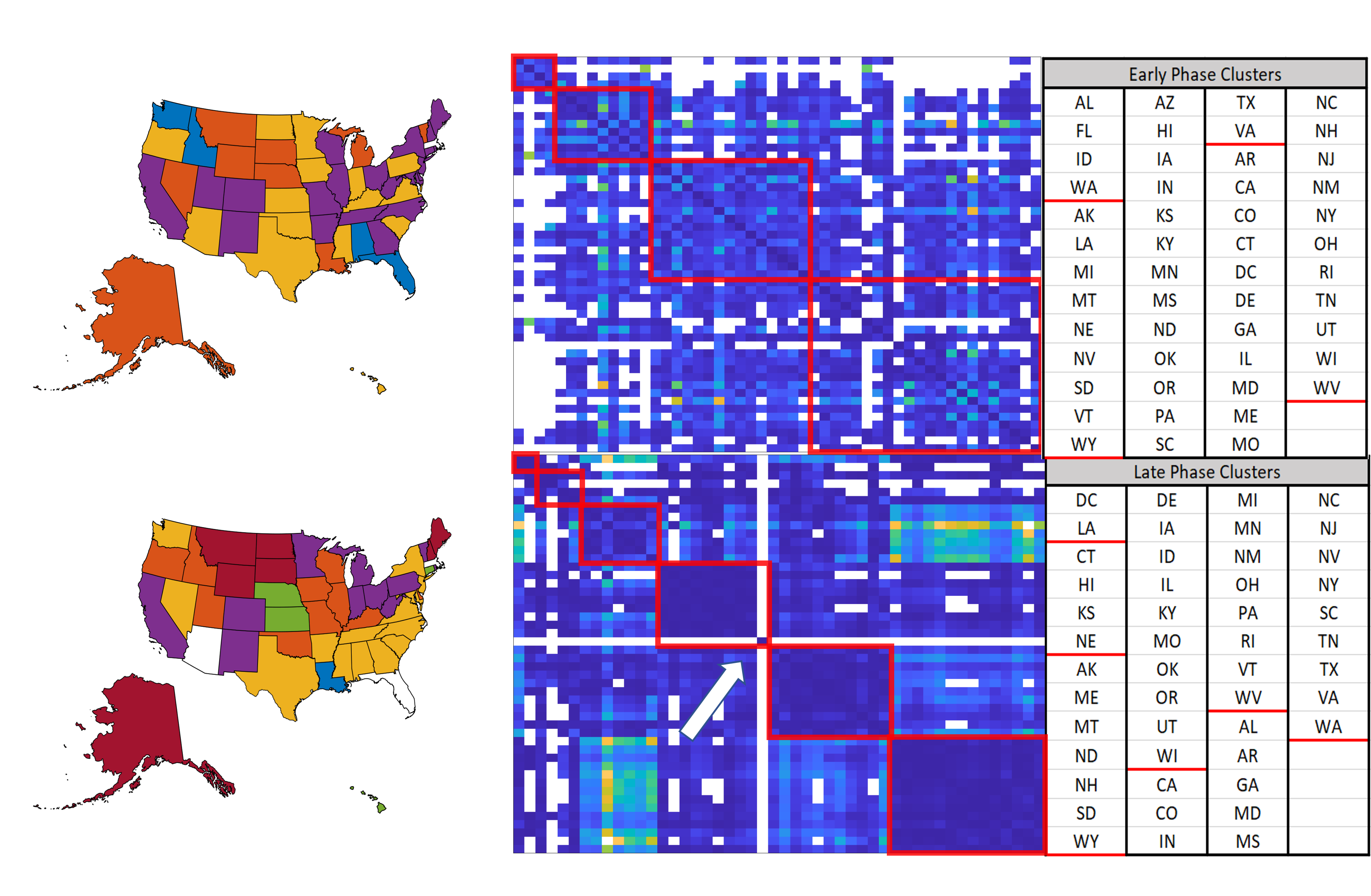}
    \caption{\label{fig:US_Map_Cluster_Figure} U.S. states are clustered in their early and late case surges via our distance metric. The dissimilarity heatmap (center top/bottom) represents the raw dissimilarity data, with white pixels indicating incompatible State curves due to incommensurate surge lengths. Red squares outline the clusters that were decided with a combination of hierarchical WPGMA clustering \cite{sokal1958statistical} and manual curation, and correspond to the listed clusters in the chart (right top/bottom). The super cluster created from manual curation in the early phase is the largest cluster, and is the bottom-right-most square in the heatmap. Massachusetts is excluded from analysis in both early and late phases due to poor data quality, and Arizona and Florida were identified as out groups for late phase clustering.}
\end{figure*}

Using methods outlined in Sections SI and SII, we first used the WPGMA algorithm \cite{sokal1958statistical} to select 14 groups of case curves from the entire corpus of data with early and late phases included. The clustering immediately separated early phase from late phase with no exceptions. 6 such clusters included early phase case curves, and 8 included late phase. Plotting the clusters, we noticed that 3 early phase clusters could be readily combined into a single super cluster. Moreover, we noticed that Wisconsin (Fig \ref{fig:US_Map_Cluster_Figure} white arrow) could be combined into another cluster, and that two clusters initially of two states (Connecticut and Hawaii, Kansas and Nebraska) may be aggregated into a single cluster of four states. This curation reduced the number of clusters to 4 early phase clusters and 6 late phase clusters.

\begin{figure*}[!ht]
    \centering
    \includegraphics[scale = .6]{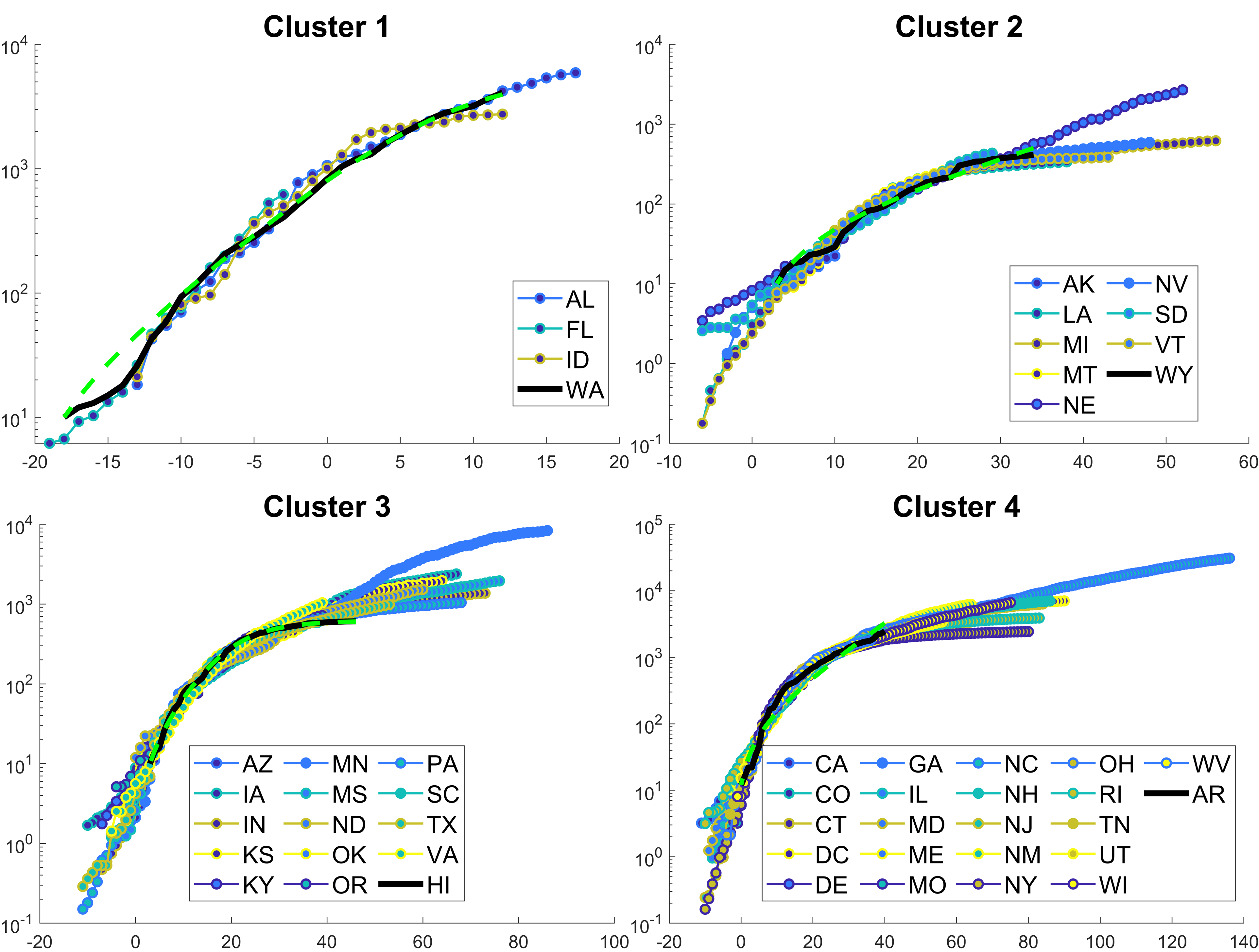}
    \caption{\label{fig:Early_Cluster} Clustered early phase cumulative case curves shown on log scale. A suitable reference curve is chosen based on average distance to other curves in its cluster, and other data is scaled and shifted to match the reference curve as closely as possible. The reference curve is named at the top of each panel and is shown as the solid black line in the plot. The SIR model fit to the reference curve is shown as a green dashed line. The horizontal axis of each plot represents time before or after the start of the reference curve's tenth case once the data have been shifted.}
\end{figure*}

Fig \ref{fig:Early_Cluster} suggests that all states in their initial case surge exhibit very similar behavior on the log scale, with minor variations towards the end of the surge as more complicated dynamics emerge. Each curve begins as a straight line and appears to be approaching level on the log scale, suggesting decelerating infection rates from an initial exponential surge in cases.

\subsection{Detecting Effective COVID-contracting Population Size}\label{sec:pop_size}

When comparing case curves metrics between states, one usually scales the number of cases by total population size in order to measure per capita statistics rather than simple raw data. However, on the relatively large U.S. State scale, normalizing by population may not be entirely appropriate, because doing so implicitly assumes that the entire population contributes to the growth in cases, and ignores the complex reality wherein cases may be localized in hot spots whose size may not correlate closely to the population as a whole. When analyzing such statistics, then, it may therefore be helpful to focus on the cities, counties, or even neighborhoods in which mixing of individuals is causing COVID transmission. By allowing population sizes to be freely rescaled when comparing states, our alignments can be used to estimate the relative sizes of COVID-impacted populations between any two given states, as mentioned in section SII. We tested whether the ratios of population size of states also predict the ratios of their COVID-contracting populations (Table \ref{tab:fit_stats}), finding that there is a weak $R^2$ of $0.31$. A similar score was seen when total State population was replaced by urban population (given by the total population of the largest 3 cities in each State (Table \ref{tab:fit_stats}, $R^2 = 0.2005$)), reflecting, perhaps the dominant role played in many states by transmission of COVID within urban populations. However, our analysis also exposed some notable exceptions. For example, our clustering ratio between Florida and the other states in its cluster were routinely lower than the actual population ratios of the states in that cluster, indicating that states matching Florida's case dynamics had to be scaled up less than the expected population ratio in order for the curves to align, suggesting that the COVID-contracting population was a smaller fraction of Florida's total population than for states with similar case dynamics. Conversely, Idaho had the opposite result (higher clustering ratio versus population ratio), indicating that it had a larger COVID-contracting population fraction than states in the same cluster.

\begin{table}
    \centering
    \begin{tabular}{m{5cm}|c}
    Correlation Tested
    & $R^2$ value\\
    \hline 
    \hline
    Clustering population ratio (log scale) vs. actual population ratio (log scale)
    & $0.3169$ \\
    \hline
    Clustering population ratio (log scale) vs. population ratio of sum of 3 largest cities (log scale)
    & $0.2005$ \\
    \hline
    \end{tabular}
    \caption{\label{tab:fit_stats} A summary of the basic linear regressions done on the data.}
\end{table}


\subsection{Searching for Variables to Explain Clustering}\label{sec:cluster_stats}

The clustering of states allows us to test hypotheses about which underlying variables, including both geographical proximity and similarities or divergences between public health measures adopted in different states. A lack of testing capacity---especially at the beginning of the COVID-19 pandemic---along with the high frequency of asymptomatic infections cloud estimates of the true prevalence of the disease. PCR test results of individuals on a cruise ship amid a COVID outbreak revealed high variability in test positivity rates between different testing methods \cite{Nagasaki_Cruise_Ship}. Antibody tests also likely depend on disease severity \cite{Hansen109}, and as a result can prove to be ineffective at retrospective analysis of COVID cases. Even so, studies that have been done using antibody tests have revealed false negatives in all methods of PCR testing \cite{Ineffective_Antibody}. Additionally, limited availability of COVID tests, particularly early in the pandemic, may mean that many individuals infected with COVID were not tested \cite{LAU2021110}. The number of positive cases is therefore likely under-counted (although estimates vary for the extent to which cases have been under-counted \cite{Ironse2103272118,case_undercount_2,LAU2021110}), creating concern that the clusters identified in this study may be distorted or even dominated by different levels of testing coverage and different kinds of testing methods. We tested whether differences in testing coverage explained the different time dependencies of cases. We used percentage of positive tests as a proxy for coverage since this indicates when testing coverage is low: for example the test positivity rate reached over 40\% in New York State during its first surge. We found that the average positivity rates within clusters were no more similar than would be expected under random grouping ($p = 0.5365$, by permutation test). We may therefore conclude that our clustering is not simply capturing differences in testing coverage.

We also performed our modified Mantel test on the data to see whether dissimilarity scores between states are correlated with physical distances between the population centroids of states being compared. Although there are few known examples of COVID hotspots spanning State boundaries, such similarity would be expected if two states are linked by high rates of migration. Some pairs of neighboring states showed expected close similarity: for example much of the Southeast, including Georgia, Alabama, and Tennessee (Fig \ref{fig:US_Map_Cluster_Figure}), as well as Idaho, North, and South Dakota. Other clusters were made up of geographically distant states: for example, Connecticut, Hawaii, Nebraska, and Kansas all emerged within a single cluster (Fig \ref{fig:US_Map_Cluster_Figure}). Accordingly, we checked whether close states tended to cluster together. The correlation between closeness of our clustering was low only $\rho = 0.0991$, but it is statistically significant ($p  = 2.5486 \times 10^{-8}$, by Mantel test, Table \ref{tab:perm_stats}). To further probe the issue, we also measured the average percentage of states in each cluster that share a bordering neighbor also in the cluster. We found that approximately 63\% of states on average had a neighbor in their cluster. This number was tested it against 50,000 permutations of our clustering for a significance of $p = 1.0868 \times 10^{-5}$. Both tests indicate that our clustering did indeed tend to cluster close states more than a random clustering would, but also that closeness is not necessarily a very important factor deciding whether dynamics are similar.

\begin{table*}
    \centering
    \begin{tabular}{m{5cm}|m{5cm}|c}
    Metric Tested 
    & Test Statistic
    & $p$ value\\
    \hline 
    \hline
    Population centroid distance
    & Pearson correlation
    & $2.5486 \times 10^{-8}$\\
    \hline
    Absolute Trump voting percentage difference
    & Pearson correlation
    & $0.1453$ \\
    \hline
    Majority party governorship status
    & Average percentage of most represented party in cluster
    & $0.3068$ \\
    \hline
    States clustered together both in early and late phase
    & Average percentage of states in each cluster that are clustered together in both early and late phase
    & $0.3893$ \\
    \hline
    States clustered with their bordering neighbors
    & Average percentage of states in each cluster that shared a neighbor in the cluster
    & $1.0868 \times 10^{-5}$\\
    \hline
    Testing positivity rates
    & Average variance of testing positivity in each cluster
    & $0.5365$\\
    \hline
    School reopening dates
    & Average variance of school reopening dates
    & $0.2516$\\
    \hline
    \end{tabular}
    \caption{\label{tab:perm_stats} A summary of results from permutation tests (including our modified Mantel tests) done on our clustering. The left column describes a metric used to probe our clustering behavior. The $p$ value in the right column refers to the percentage of time that the metric tested on a permuted version of our clustering exceeded that of our actual clustering. For each test, the number of permutations tested is 50,000.}
\end{table*}


Public health responses within states are strongly influenced by political climate and divisions between the responses of the two main political parties to a joint public health and economic crisis. We considered two measures of political climate: the party affiliation of the State governor at the beginning of the pandemic (excluding Washington D.C., which does not have a governor), and the percentage of voters who voted for the Republican candidate, Donald Trump, in the 2020 presidential elections. Neither measure was significantly correlated with State clustering (governor party, $p = 0.3068$, and Trump voting percentage, $p = 0.1453$).

Finally, we use school closures as a direct index of the level of social distancing enforced in each State. Some states (e.g. Florida and Rhode Island) adopted a single policy on keeping schools opened or closed; in other states, individual school districts or counties determined whether or not to offer in-person instruction. Hence, we used the largest school district in each State as a proxy for the entire State's response to the pandemic. We compiled data on when the largest school district in each State first opened some form of in-person learning, and found that whether or not schools were open during the last recorded surge did not correlate with our State clustering ($p = 0.2516$).

\section{Discussion}

Comparing COVID-19 cases between populations: neighborhoods, cities, states, countries, is an important part of building narratives for understanding the impacts of the disease, and the roles of policy, inequality and racism in shaping its severity. But case numbers can not be compared without correcting for differences in population sizes. Normalizing by the size of the population can lead to misleading comparisons, if COVID-19 incidence varies in systematic ways among individuals in the population. Here we sought to make comparisons between populations (states), under the simplifying assumption that each contains an effectively well-mixed subpopulation of COVID-19 contacting individuals. Importantly, detecting the size of this population is an element in our algorithm. 

Clustering of data from different states illuminates a second, general, question, that will be explored more deeply in Part II of this work. Specifically, although modern tools in epidemic modeling can incorporate population-heterogeneity, because of their many parameters, these models are challenging to fit to real data from the COVID-19 pandemic.  The assumption that cases are dominated by transmission within a subpopulation of COVID-19 contacting individuals, who are effectively well-mixed, motivates us to further consider that there may be a limited number of archetypal case growth curves among this subpopulation. We find through clustering, that when properly normalized, and in spite of very different geographies and demographic profiles, cases in different states follow one of remarkably few (4-6) growth archetypes. 

Some of the conclusions that follow from the clustering are expected---for example we found that the late stage of the pandemic behaves distinctly from the early stage, as indicated by the fact that the clustering algorithm separated the two phases without direction. Due to the wildly different social and epidemiological conditions between the two time periods, this is unsurprising. 

The first phase conforms extremely closely to simple SIR dynamics, as indicated by Fig \ref{fig:Early_Cluster}, suggesting a simple and effective mean-field description of the early dynamics. The fits shown in the figure come from simple model smoothing: we found appropriate constant transmission rate $\beta$ and population size $N$ values for the SIR model that best fit the data in a least-squares sense (see Supplementary Materials section SVI). Importantly, fitting $N$ to the data enables the entire first wave to be fit with constant disease transmission parameters, and thus with a constant basic reproductive rate $R_0$ \cite{diekmann1990definition}. In addition to their greater simplicity, the constant $\beta$, $N$ models draw a very different conclusion about how the wave ends than a time-varying $R_0$ model---for example the fits for Washington (shown in Fig \ref{fig:Early_Cluster}) give an $R_0$ value of $1.14$ within an affected population of about $27000$ . At this value of $R_0$, the disease spreads through its affected population, and case numbers only start to decrease once $1-1/R_0 \approx 12\%$ of the affected population has acquired resistance to the disease. The actual downturn occurred well before this 12\%. The fits therefore are consistent with almost immediate success of shelter-in-place measures in preventing the disease from spreading beyond a fixed number of closely-contacting individuals, but continued, unchecked spread of the disease within this group. This is in contrast to other, more complex methods of data fitting that relied on a time-variable reproduction number $R$ \cite{Rtlive,THOMPSON2019100356,Cori_Rt,sam_abbott_2021_5036949}. Hence, although actual transmission behavior is certainly heterogeneous and occurs in multiple subpopulations, a simple, constant, bulk parameter suffices to describe early epidemic dynamics within populations with a large percentage of susceptible individuals, as explored further in part II of this paper.

\begin{figure*}[!ht]
    \centering
    \includegraphics[scale = .4]{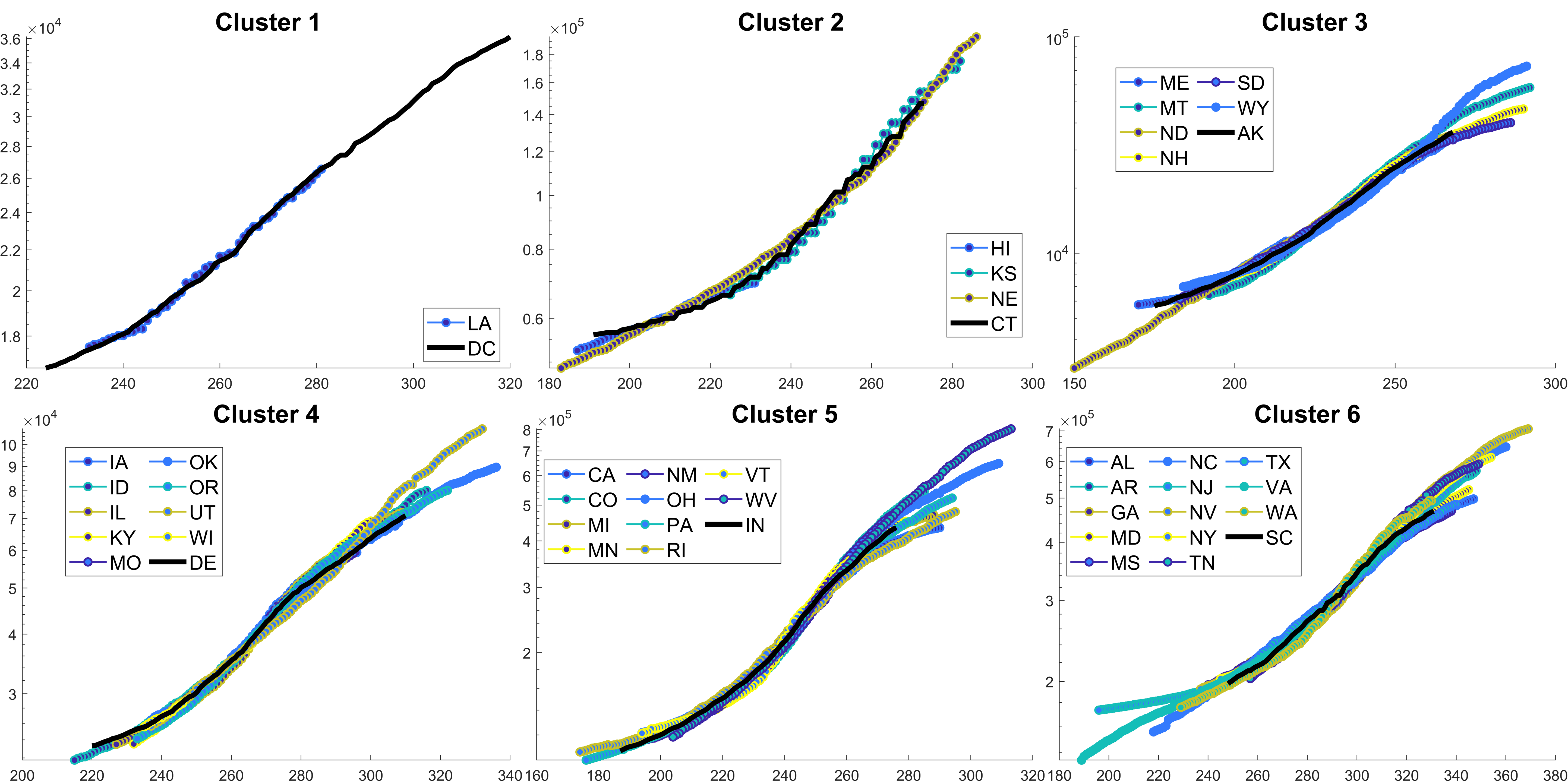}
    \caption{\label{fig:Late_Cluster} Clustered late phase cumulative case curves are shown on the log scale. A suitable reference curve is chosen based on average distance to other curves in its cluster, and other data is scaled and shifted to match the reference curve as closely as possible. The reference curve is named at the top of each panel and is shown as the solid black line in the plot. The horizontal axis of each plot represents time before or after the start of the reference curve's tenth case once the data have been shifted.}
\end{figure*}

What reveals itself unexpectedly is just how similar the late stage dynamics are from cluster to cluster. There are three factors separating late stage surge dynamics of cumulative cases on the log scale indicated by Fig \ref{fig:Late_Cluster}: One; the overall rate of increase, i.e. the overall ``slope" of the curve; Two; the degree to which the ``S" curve deviates from the straight line; Three; the time scale over which the curve displays its ``S." A possible fourth factor is the precise position of the inflection point of the ``S" curve. This is significant since, despite wildly varying conditions between and within states, there seems to exist only three or four bulk parameters which suffice to describe the spread of COVID during a surge. This leads to a reassuring discovery: during surges, the simplest models suffice. At the beginning, the SIR model does well to describe dynamics, and later, some new but equally simple model can likely be found.

We also observe that a pair of states sharing behavior at an early phase is not predictive of the same pair sharing behavior later. This can be seen by examining the totally different cluster structure between the two phases, and is confirmed empirically using a permutation test. We measured the number of states sharing the same cluster, concluding that 37 unique pairs of states shared a cluster in both the early phase and the late phase (Texas and South Carolina, for example). This number is not significantly high as indicated by permutation test (approximately 40\% of statistics from the permuted data were higher, see Table \ref{tab:perm_stats}), indicating what we see at a glance: behavior shared early does not imply behavior shared late. Comparisons which were important towards the beginning of the pandemic may not necessarily be appropriate at later times, further supporting our comparison of entire time courses.

\begin{figure*}[!ht]
    \centering
    \includegraphics[scale = .6]{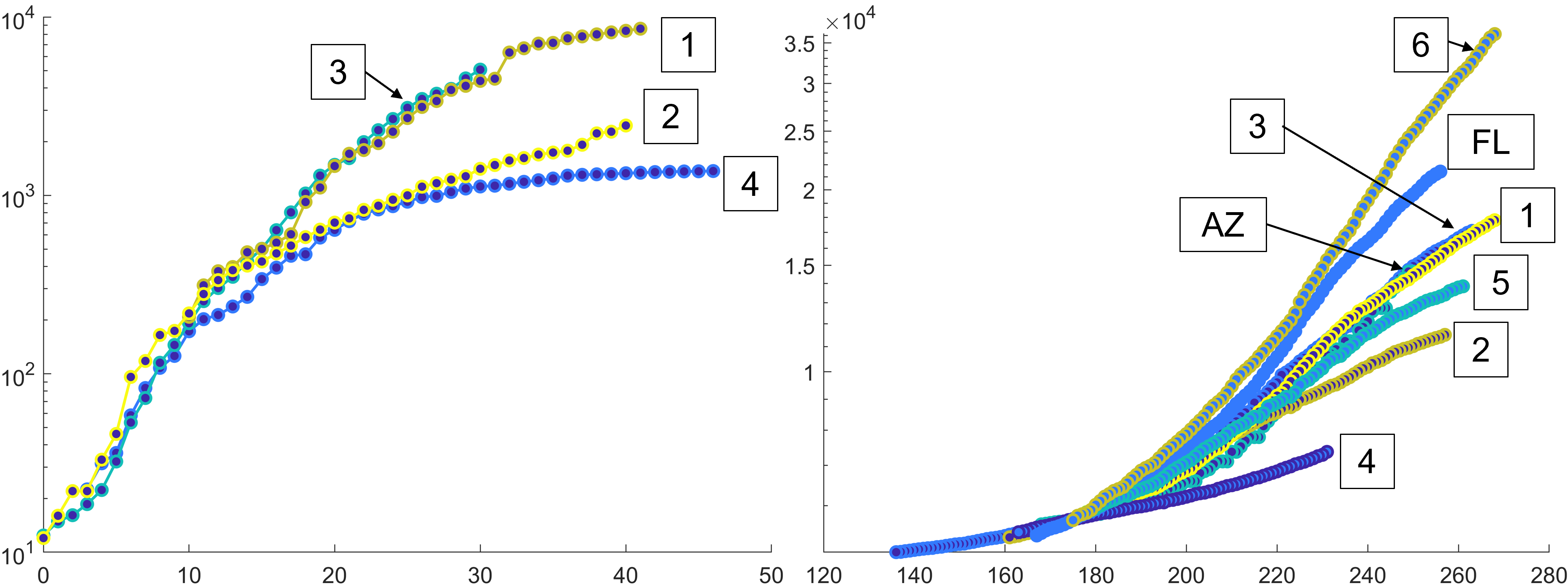}
    \caption{\label{fig:Cross_Cluster} Representative curves from each respective cluster are compared with alignment biased toward the beginning of the curve in order to emphasize disparate evolution of dynamics, shown on log scale. Number labels correspond to cluster IDs, and match those shown in Figures \ref{fig:Early_Cluster} and \ref{fig:Late_Cluster}. Left: Early phase case curves. Right: Late phase case curves, including Florida and Arizona, which are excluded from previous analyses.}
\end{figure*}

Our clustering method provides a data-driven, nonparametric way of classifying and comparing State-by-State COVID dynamics. This new approach revealed that comparisons between states using metrics that are devised \textit{a priori} are perhaps misleading. For instance, California and Florida have often been compared \cite{Tampa_Comparison,CAFLTX_Compare,Healthline_Compare} since these are two states boasting large populations with multiple urban centers and warm climates, yet having very different responses to COVID---for example, during the second surge, both of the largest school districts in California (Los Angeles and San Diego Unified) were closed to in-person instruction, while in Florida a State order required that all schools offer in-person instruction after July 2020. Upon Florida's reopening in August, per capita new cases were 20 new cases per 100,000 in California and 45 new cases per 100,000 in Florida. However, our analysis emphasizes that heterogeneities in COVID case intensities across each State make the total State population a weak normalizing factor when comparing different states. When normalized by the detected size of the COVID-affected population we find that California's cases grew in line with Colorado and New Mexico, but most surprisingly also in line with Indiana, Michigan, and Vermont. Conversely, in the late stage Florida's dynamics (and Arizona's) distinguished themselves from the rest of the states, including California, mostly due to a wildly different ``slope" of the general late-stage increase on the log scale, and partly due to a much earlier and drawn-out surge (Fig \ref{fig:Cross_Cluster}) (Florida's surge began in late October, and other states' tended to begin in late November, coinciding with Thanksgiving). This singles out Florida as unique from every State, not just California, and tempers any conclusions that can be drawn from the simple two-way comparison. Our results indicate that California, whose late behavior is represented in Fig \ref{fig:Cross_Cluster} cluster 5, experiences a gentler exponential growth rate with a more pronounced downturn that happens earlier (relative to the late surge) than any downturn in Florida's dynamics. In particular, differences in curve starts and ends will cause instantaneous metrics to give misleading comparisons between the two states.

Among the variables tested, only proximity of states showed any statistically significant correlation with the clustering devised. This in itself is an important result when evaluating the probable effect of measures that may carry other public health costs, such as school closures \cite{NBERw28264}. While the cluster a state belongs to controls the trajectory of the case numbers, the normalization of these numbers, which we have interpreted to be the number of COVID-contracting individuals, may be yet influenced by these or other variables, and independent evaluation of the population sizes, and whether they can be predicted from population, geographic, political, or public heath data is certainly warranted.

If the eventual goal in COVID data comparisons is to decide which health measures, public behaviors, and demographies lend themselves best to weathering a pandemic, then we must take care to compare the right data at the right times. Our analysis highlights two features. First, we show that pairwise comparisons of states can only be made in the context of a per-population case number that indicates the number of COVID-affected individuals, and not the total population of the State. Second, once the normalization is found, states conform to one of a small number of archetypal case curves. For early phase of COVID we can explicitly fit these case curves by an SIR model; in the latter phase, the well-mixed model no longer fits the data, yet the robust similarity of case curves suggests that models with comparably few degrees of freedom may yet be used to fit the data. Given the abundance of pandemic data now available, we have the tools to examine entire time courses at different scales (e.g. city or country) within the U.S. and internationally and compare in context.

\bibliography{bibliography}

\end{document}